\documentclass[twocolumn,aps,prd,floatfix,nofootinbib,superscriptaddress]{revtex4}
\usepackage{epsfig}
\usepackage{graphicx}
\usepackage{amssymb,amsmath}
\usepackage{pdflscape}

\newcommand{\Obig}[1]{{\mathcal{O}\left (#1\right )}}

\begin{document}

\title{\bf 
Measurement of the $e^+e^-\to\eta\gamma$ cross section near the $\phi(1020)$ resonance\\
with the SND detector}

\author{M.~N.~Achasov} 
\affiliation{Budker Institute of Nuclear Physics, SB RAS, Novosibirsk, 630090, Russia} 
\affiliation{Novosibirsk State University, Novosibirsk, 630090, Russia} 
\author{A.~E.~Alizzi}
\affiliation{Budker Institute of Nuclear Physics, SB RAS, Novosibirsk, 630090, Russia}
\affiliation{Novosibirsk State University, Novosibirsk, 630090, Russia}
\author{A.~Yu.~Barnyakov} 
\affiliation{Budker Institute of Nuclear Physics, SB RAS, Novosibirsk, 630090, Russia}
\affiliation{Novosibirsk State University, Novosibirsk, 630090, Russia} 
\author{E.~V.~Bedarev} 
\affiliation{Budker Institute of Nuclear Physics, SB RAS, Novosibirsk, 630090, Russia} 
\affiliation{Novosibirsk State University, Novosibirsk, 630090, Russia} 
\author{K.~I.~Beloborodov}
\affiliation{Budker Institute of Nuclear Physics, SB RAS, Novosibirsk, 630090, Russia} 
\affiliation{Novosibirsk State University, Novosibirsk, 630090, Russia} 
\author{A.~V.~Berdyugin} 
\affiliation{Budker Institute of Nuclear Physics, SB RAS, Novosibirsk, 630090, Russia} 
\affiliation{Novosibirsk State University, Novosibirsk, 630090, Russia} 
\author{A.~G.~Bogdanchikov} 
\affiliation{Budker Institute of Nuclear Physics, SB RAS, Novosibirsk, 630090, Russia} 
\author{A.~A.~Botov} 
\affiliation{Budker Institute of Nuclear Physics, SB RAS, Novosibirsk, 630090, Russia} 
\author{D.~E.~Chistyakov}
\affiliation{Budker Institute of Nuclear Physics, SB RAS, Novosibirsk, 630090, Russia} 
\affiliation{Novosibirsk State University, Novosibirsk, 630090, Russia} 
\author{T.~V.~Dimova} 
\affiliation{Budker Institute of Nuclear Physics, SB RAS, Novosibirsk, 630090, Russia} 
\affiliation{Novosibirsk State University, Novosibirsk, 630090, Russia} 
\author{V.~P.~Druzhinin} 
\affiliation{Budker Institute of Nuclear Physics, SB RAS, Novosibirsk, 630090, Russia} 
\affiliation{Novosibirsk State University, Novosibirsk, 630090, Russia} 
\author{L.~V.~Kardapoltsev}
\email{l.v.kardapoltsev@inp.nsk.su}
\affiliation{Budker Institute of Nuclear Physics, SB RAS, Novosibirsk, 630090, Russia} 
\affiliation{Novosibirsk State University, Novosibirsk, 630090, Russia} 
\author{A.~S.~Kasaev} 
\affiliation{Budker Institute of Nuclear Physics, SB RAS, Novosibirsk, 630090, Russia} 
\author{A.~A.~Kattsin}
\affiliation{Budker Institute of Nuclear Physics, SB RAS, Novosibirsk, 630090, Russia} 
\author{A.~G.~Kharlamov} 
\affiliation{Budker Institute of Nuclear Physics, SB RAS, Novosibirsk, 630090, Russia} 
\affiliation{Novosibirsk State University, Novosibirsk, 630090, Russia} 
\author{I.~A.~Koop}
\affiliation{Budker Institute of Nuclear Physics, SB RAS, Novosibirsk, 630090, Russia} 
\affiliation{Novosibirsk State University, Novosibirsk, 630090, Russia} 
\author{A.~A.~Korol} 
\affiliation{Budker Institute of Nuclear Physics, SB RAS, Novosibirsk, 630090, Russia} 
\affiliation{Novosibirsk State University, Novosibirsk, 630090, Russia} 
\author{D.~P.~Kovrizhin} 
\affiliation{Budker Institute of Nuclear Physics, SB RAS, Novosibirsk, 630090, Russia} 
\author{A.~S.~Kupich} 
\affiliation{Budker Institute of Nuclear Physics, SB RAS, Novosibirsk, 630090, Russia} 
\affiliation{Novosibirsk State University, Novosibirsk, 630090, Russia} 
\author{A.~P.~Kryukov} 
\affiliation{Budker Institute of Nuclear Physics, SB RAS, Novosibirsk, 630090, Russia} 
\author{N.~A.~Melnikova} 
\affiliation{Budker Institute of Nuclear Physics, SB RAS, Novosibirsk, 630090, Russia} 
\author{N.~Yu.~Muchnoi} 
\affiliation{Budker Institute of Nuclear Physics, SB RAS, Novosibirsk, 630090, Russia} 
\affiliation{Novosibirsk State University, Novosibirsk, 630090, Russia} 
\author{A.~E.~Obrazovsky} 
\affiliation{Budker Institute of Nuclear Physics, SB RAS, Novosibirsk, 630090, Russia} 
\author{A.~A.~Oorzhak}
\affiliation{Budker Institute of Nuclear Physics, SB RAS, Novosibirsk, 630090, Russia}
\affiliation{Novosibirsk State University, Novosibirsk, 630090, Russia}
\author{E.~V.~Pakhtusova} 
\affiliation{Budker Institute of Nuclear Physics, SB RAS, Novosibirsk, 630090, Russia} 
\author{I.~A.~Polomoshnov}
\affiliation{Budker Institute of Nuclear Physics, SB RAS, Novosibirsk, 630090, Russia}
\affiliation{Novosibirsk State University, Novosibirsk, 630090, Russia}
\author{K.~V.~Pugachev} 
\affiliation{Budker Institute of Nuclear Physics, SB RAS, Novosibirsk, 630090, Russia} 
\affiliation{Novosibirsk State University, Novosibirsk, 630090, Russia} 
\author{S.~A.~Rastigeev} 
\affiliation{Budker Institute of Nuclear Physics, SB RAS, Novosibirsk, 630090, Russia} 
\author{Yu.~A.~Rogovsky} 
\affiliation{Budker Institute of Nuclear Physics, SB RAS, Novosibirsk, 630090, Russia} 
\affiliation{Novosibirsk State University, Novosibirsk, 630090, Russia} 
\author{A.~I.~Senchenko} 
\affiliation{Budker Institute of Nuclear Physics, SB RAS, Novosibirsk, 630090, Russia} 
\author{S.~I.~Serednyakov} 
\affiliation{Budker Institute of Nuclear Physics, SB RAS, Novosibirsk, 630090, Russia} 
\affiliation{Novosibirsk State University, Novosibirsk, 630090, Russia} 
\author{Yu.~M.~Shatunov} 
\affiliation{Budker Institute of Nuclear Physics, SB RAS, Novosibirsk, 630090, Russia} 
\author{D.~A.~Shtol} 
\affiliation{Budker Institute of Nuclear Physics, SB RAS, Novosibirsk, 630090, Russia} 
\author{Z.~K.~Silagadze} 
\affiliation{Budker Institute of Nuclear Physics, SB RAS, Novosibirsk, 630090, Russia} 
\affiliation{Novosibirsk State University, Novosibirsk, 630090, Russia} 
\author{K.~D.~Sungurov}
\email{k.d.sungurov@inp.nsk.su}
\affiliation{Budker Institute of Nuclear Physics, SB RAS, Novosibirsk, 630090, Russia} 
\affiliation{Novosibirsk State University, Novosibirsk, 630090, Russia} 
\author{I.~K.~Surin} 
\affiliation{Budker Institute of Nuclear Physics, SB RAS, Novosibirsk, 630090, Russia} 
\author{Yu.~V.~Usov} 
\affiliation{Budker Institute of Nuclear Physics, SB RAS, Novosibirsk, 630090, Russia} 
\author{V.~N.~Zhabin} 
\affiliation{Budker Institute of Nuclear Physics, SB RAS, Novosibirsk, 630090, Russia} 
\affiliation{Novosibirsk State University, Novosibirsk, 630090, Russia} 
\author{Yu.~M.~Zharinov} 
\affiliation{Budker Institute of Nuclear Physics, SB RAS, Novosibirsk, 630090, Russia}
\author{V.~V.~Zhulanov}
\affiliation{Budker Institute of Nuclear Physics, SB RAS, Novosibirsk, 630090, Russia}
\affiliation{Novosibirsk State University, Novosibirsk, 630090, Russia}

\begin{abstract} 
In the experiment with the SND detector at the VEPP-2000 $e^+e^-$ collider, the $e^+e^-\to\eta\gamma$ 
cross section is measured in the energy range $E = 980 - 1060$ MeV.
The measurement is carried out in the $\eta\to 2\gamma$ decay mode.
Data with an integrated luminosity of 73 pb$^{-1}$ collected in 2018 and 2024 are used in the analysis. 
The measured cross section is the most accurate to
date and is higher compared to other measurements.
From the fit to the cross section data with the vector meson dominance model, 
the value of the product of the branching fractions
$B(\phi\to e^+e^-)B(\phi\to \eta\gamma)$ has been obtained.
\end{abstract}

\pacs{13.66.Bc, 14.40.Aq, 13.40.Gp}

\maketitle

\section{Introduction}

Radiative decays of the light vector mesons $\rho$, $\omega$, $\phi$ and particularly
the decay of the $\phi$ meson to $\eta \gamma$ are known as a good laboratory
for testing various theoretical concepts describing the low-energy QCD dynamics. Their measured decay widths 
are widely used to test predictions based on the quark model~\cite{quarkmodel1,quarkmodel2} 
and study the $SU(3)$ flavor symmetry breaking~\cite{BHLS}.
They are also used to study the 
$\eta-\eta^{\prime}$ mixing and the possible admixture of an additional $SU(3)$ singlet state~\cite{etamix1,etamix2,etamix3},
which can be attributed to the pseudoscalar glueball predicted by QCD~\cite{glue} 
and to the $c\bar{c}$ state~\cite{ccbar}.

This work is devoted to measurement of the $e^+e^-\to\eta\gamma$ cross section
in the center-of-mass (c.m.) energy range $E = 980 - 1060$ MeV using data collected by
the Spherical Neutral Detector (SND) at the $e^+e^-$ collider VEPP-2000. In this energy range,
the dominant contribution to the cross section comes from the decay of $\phi\to\eta\gamma$,
which interferes with the tails of $\rho(770)$ and high-lying resonances. The $\phi\to\eta\gamma$ 
branching fraction can be obtained from fitting the $e^+e^-\to\eta\gamma$ cross section within
the vector meson dominance (VMD). Near the $\phi(1020)$ resonance, the most accurate measurements 
of the $e^+e^-\to\eta\gamma$ cross section
were made by the SND~\cite{SNDetg2000, SNDetg2007} and 
CMD-2~\cite{CMDetg1999, CMDetg2001, CMDetg2005} experiments at the VEPP-2M collider.

\section{SND detector}
The SND detector\cite{SNDdet1,SNDdet2,SNDdet3,SNDdet4} is a general-purpose non-magnetic detector 
operating at the VEPP-2000 $e^+e^-$ collider~\cite{vepp2000} with the c.m. energy $E = 0.32 -
2$ GeV. The main part of the SND detector 
is an electromagnetic calorimeter. The SND calorimeter is made of 1640 
NaI(Tl) crystals. It consists of three spherical layers 
with the total thickness of 13.4 radiation lengths. The calorimeter covers 
nearly $95\%$ of the full solid angle: $18^\circ\le\theta\le 162^\circ $, 
where $\theta $ is the polar angle. The energy resolution of the SND
calorimeter is given by the formula 
$\sigma_E/E _{\gamma}(\%)=4.2\%/\sqrt[4]{E_{\gamma}(\mbox{GeV})}$, 
while the angular resolution is 
$\sigma_{\varphi}=0.82^\circ/\sqrt{E _{\gamma}(\mbox {GeV})}\oplus 
0.63^\circ$. The parameters of charged particles are measured 
using a nine-layer drift chamber and a proportional chamber with
cathode-strip readout located in a common gas volume.
Its angular resolution is $0.8^\circ$ in polar 
direction, and $0.45^\circ$ in the azimuthal direction.
The solid angle of the tracking system is 94\% of $4\pi$.
A system of aerogel threshold Cherenkov counters is located around the tracking system.
The calorimeter 
is surrounded by a 10 cm thick iron absorber, behind which there is 
a muon system consisting of proportional tubes and scintillation counters.

For this work we use data 
collected by the SND detector at the $e^+e^-$ collider VEPP-2000 in the 
c.m. energy range $E = 980 - 1060$ MeV. The integrated luminosity
accumulated at 21 energy points in 2024 and at 8 energy points in 2018 
is about 73 pb$^{-1}$.

Monte-Carlo (MC) simulation of signal and background processes takes into account 
radiative corrections~\cite{radcor}. The angular distribution of a hard photon emitted from 
the initial state is generated according to Ref.~\cite{BM}. Interactions of the particles 
produced in $e^{+}e^{-}$ annihilation with the detector materials are simulated using 
the GEANT4 software~\cite{geant}. The simulation takes into account variation of experimental 
conditions during data taking, in particular dead detector channels and beam-induced background. 
To take into account the effect of superimposing the beam background on
the $e^{+}e^{-}$ annihilation events, the simulation uses special background events recorded 
during data taking with a random trigger. These events are superimposed on the simulated events, 
leading to the appearance of additional tracks and photons in them.

The average beam energy and the energy spread were measured during data taking
by a dedicated system using the Compton backscattering of laser photons on the electron beam~\cite{compton}.
Energy measurements performed at a given energy point are averaged 
with weights proportional to the integrated luminosity. The obtained 
values for average energy, energy spread $\sigma_{E}$ and their errors are listed in Table~\ref{tab1}. 

The energy errors listed in Table~\ref{tab1} describe the relative uncertainties of 
the measured energy. Additional uncertainty comes
from a possible shift of the energy scale. 
It  was estimated to be 60 keV at the c.m. energy $E $ = 1000 MeV
by comparison with the energy measurement by the resonance depolarization method~\cite{compton}.
Since the measurement of the $\phi$
meson mass is not a subject of this work, we only need to know the value
of this shift relative to the $\phi$ meson mass ($\Delta m_{\phi}$).
The relative shift of the energy scale for the 2018 
energy scan was obtained in Ref.~\cite{SNDkskl} by fitting the energy dependence 
of the $e^+e^-\to K_SK_L$ cross section. 
The measured energies used in Ref.~\cite{SNDkskl} differ slightly
from the ones used in this work due to slightly different calibration procedures 
applied to the data. With the set of measured energies used in this work,
the fit of the $e^+e^-\to K_SK_L$ cross section described in Ref.~\cite{SNDkskl} gives 
a common shift of all energy points in the 2018 scan $\Delta m_{\phi} = 0.020 \pm 0.012$ MeV.
Taking into account the interaction of K mesons in final state changes it 
to $\Delta m^{FSI}_{\phi} = 0.035 \pm 0.012$ MeV. When fitting the energy dependence 
of the $e^+e^-\to\eta\gamma$ cross section, we introduce a common energy shift with 
a Gaussian constraint corresponding to $\Delta m^{FSI}_{\phi}$. The energy shift  
$\Delta m_{\phi}$ is used to estimate the associated systematic uncertainty.

We use the $e^+e^-\to\gamma\gamma$ process to measure the luminosity.
For event selection, determination of systematic uncertainties and luminosity 
corrections, we follow the procedure developed in Ref.~\cite{SNDkskl}.
For the coherence with the selection conditions of the $e^+e^-\to\eta\gamma$ events, we additionally
impose selection conditions on the normalized energy deposition in the calorimeter $E_{tot}/E > 0.65$
and on the normalized transverse momentum of the event calculated using the energy 
depositions in the calorimeter crystals $P_{t}/E < 0.3$. 

The process under study $e^+e^-\to\eta\gamma\to 3\gamma$ contains only photons
in the final state. The choice of the $e^+e^-\to\gamma\gamma$ process for the luminosity measurement
allows us to cancel systematic uncertainties in the measured cross section that are
common to both of these processes.
The most important of them are the systematic uncertainties associated with 
the simulation of the hardware first-level trigger and the superimposition of beam-generated spurious
tracks onto signal events. For this reason, we do not apply the correction on the superimposition of 
beam-generated tracks introduced in Ref.~\cite{SNDkskl}. 

The resulted integrated luminosity for 
each energy point is presented in Table~\ref{tab1} with statistical and systematic errors. 
The systematic uncertainty does not exceed 0.6\%.

\begin{table}
\caption{\label{tab1} The center-of-mass energy (E), center-of-mass energy 
spread ($\sigma_{E}$), and integrated luminosity (IL). For luminosity,
the first error is statistical, the second is systematic.}
\begin{ruledtabular}
\begin{tabular}{ccc}
E, MeV & $\sigma_{E}$, keV & IL, nb$^{-1}$\\
\hline
\multicolumn{3}{c}{2024 energy scan} \\
980.189 $\pm$ 0.010 & 470 $\pm$ 97 & 2061.3 $\pm$ 4.1 $\pm$ 8.9 \\
984.124 $\pm$ 0.012 & 543 $\pm$ 120& 2205.9 $\pm$ 4.2 $\pm$ 7.7 \\
990.032 $\pm$ 0.009 & 439 $\pm$ 77 & 1914.7 $\pm$ 4.0 $\pm$ 9.4 \\
1000.039 $\pm$ 0.018 & 340 $\pm$ 108 & 4038.3 $\pm$ 5.8 $\pm$ 15.9 \\
1002.038 $\pm$ 0.015 & 371 $\pm$ 128 & 1988.8 $\pm$ 4.1 $\pm$ 8.3 \\
1006.034 $\pm$ 0.026 & 435 $\pm$ 160 & 2297.2 $\pm$ 4.4 $\pm$ 5.4 \\
1010.091 $\pm$ 0.014 & 462 $\pm$ 74  & 2161.9 $\pm$ 4.3 $\pm$ 8.3 \\
1013.099 $\pm$ 0.008 & 449 $\pm$ 53 & 2084.7 $\pm$ 4.2 $\pm$ 9.5 \\
1016.110 $\pm$ 0.010 & 444 $\pm$ 69 & 3113.8 $\pm$ 5.2 $\pm$ 10.1 \\
1017.073 $\pm$ 0.009 & 439 $\pm$ 54 & 3354.9 $\pm$ 5.4 $\pm$ 13.2 \\
1018.111 $\pm$ 0.005 & 465 $\pm$ 39 & 3539.2 $\pm$ 5.5 $\pm$ 11.7 \\
1019.073 $\pm$ 0.004 & 444 $\pm$ 38 & 4724.9 $\pm$ 6.4 $\pm$ 12.2 \\
1020.055 $\pm$ 0.004 & 429 $\pm$ 40 & 4734.9 $\pm$ 6.4 $\pm$ 22.8 \\
1021.005 $\pm$ 0.005 & 449 $\pm$ 36 & 3607.3 $\pm$ 5.6 $\pm$ 10.4 \\
1023.079 $\pm$ 0.005 & 447 $\pm$ 43 & 3437.3 $\pm$ 5.5 $\pm$ 21.0 \\
1025.003 $\pm$ 0.008 & 355 $\pm$ 80 & 2025.1 $\pm$ 4.2 $\pm$ 7.4 \\
1025.058 $\pm$ 0.005 & 429 $\pm$ 48 & 2217.0 $\pm$ 4.4 $\pm$ 11.7 \\
1030.072 $\pm$ 0.006 & 409 $\pm$ 44 & 2577.1 $\pm$ 4.8 $\pm$ 7.9 \\
1040.047 $\pm$ 0.009 & 412 $\pm$ 48 & 2094.0 $\pm$ 4.4 $\pm$ 14.1 \\
1049.992 $\pm$ 0.009 & 438 $\pm$ 107& 4895.1 $\pm$ 6.8 $\pm$ 22.0 \\
1060.206 $\pm$ 0.012 & 444 $\pm$ 56 & 599.9 $\pm$ 2.4 $\pm$ 3.6 \\
\multicolumn{3}{c}{2018 energy scan} \\
1005.999 $\pm$ 0.021 & 359 $\pm$ 39 & 1681.6 $\pm$ 3.8 $\pm$ 4.8 \\
1016.843 $\pm$ 0.023 & 351 $\pm$ 64 & 1650.1 $\pm$ 3.7 $\pm$ 3.0 \\
1017.959 $\pm$ 0.023 & 375 $\pm$ 84 & 1256.4 $\pm$ 3.3 $\pm$ 3.5 \\
1019.101 $\pm$ 0.019 & 380 $\pm$ 36 & 2454.3 $\pm$ 4.6 $\pm$ 7.2 \\
1019.982 $\pm$ 0.019 & 380 $\pm$ 61 & 2640.2 $\pm$ 4.8 $\pm$ 9.7 \\
1020.932 $\pm$ 0.046 & 406 $\pm$ 26 & 1424.4 $\pm$ 3.5 $\pm$ 3.0 \\
1022.114 $\pm$ 0.030 & 372 $\pm$ 36 & 1229.1 $\pm$ 3.2 $\pm$ 4.8 \\
1022.929 $\pm$ 0.041 & 382 $\pm$ 43 & 821.3 $\pm$ 2.7 $\pm$ 1.2 \\ 
\end{tabular}
\end{ruledtabular}
\end{table}

\section{Selection of $e^+e^-\to\eta\gamma$ events \label {evsel}}
The selection of $e^+e^-\to\eta\gamma$ events is performed in two stages.
At the first stage, we select events with at least 3 photons with 
energies above 20 MeV, without charged tracks, $E_{tot}/E > 0.65$ and $P_{t}/E < 0.3$. 
For events passing these selection conditions, we perform a kinematic fit to the
$e^+e^-\to3\gamma$ hypotheses with the requirements of 
energy and momentum conservation.
For kinematic fit, we choose only those photons in the event which
have energy $E_{\gamma} > 50$ MeV and polar angle $27^{\circ} < \theta_{\gamma} < 153^{\circ}$.
If the number of photons in an event that meets these conditions is less than 3 or more than 8,
then such event is discarded. 
As a result of the kinematic fit, the energies and angles of photons are refined, 
and the $\chi^2$ of the $e^+e^-\to3\gamma$ kinematic hypothesis is calculated ($\chi^2_{3\gamma}$).
In the kinematics fitting, all possible combinations of photons are tested, and 
the combination with the smallest $\chi^2$ value is retained.
For further analysis, we use the refined values of photon angles and energies,
and impose the following selection condition on obtained $\chi^2$ of 
the kinematic fit $\chi^2_{3\gamma} < 30$.

To reconstruct the invariant mass of two photons from the $\eta$ meson decay ($M_{\gamma\gamma}$), 
we employ the procedure described in Ref.~\cite{CMDetg2005}. The photon energies refined after
kinematic fit are sorted in descending order $E_1 > E_2 > E_3$. If $E_3 < m_{\eta}^{2}/E$, then
$M_{\gamma\gamma} = \sqrt{4E_{b}(E_b-E_{2})}$, otherwise $M_{\gamma\gamma} = \sqrt{4E_{b}(E_b-E_{1})}$.
Here $m_{\eta}$ is the $\eta$ meson mass and $E_b = E/2$. 
The following constraint is imposed on the obtained invariant mass $350 < M_{\gamma\gamma} < 700$ MeV.

The dominant background for this analysis arises from the QED process
of $e^+e^-$ annihilation to 3 photons. For an accurate description of the background,
it is also important to take into account the annihilation into 2 photons with an additional photon from the splitting of 
the electromagnetic shower or beam background, and the annihilation into four and more photons.
To model these processes, we generate a sample of the $e^+e^- \to \gamma \gamma (\gamma)$
events using the Monte Carlo event generator BabaYaga@NLO~\cite{BabaYaga}. For $e^+e^- \to 2\gamma, 3\gamma$,
it exploits the exact amplitude up to the $\Obig{\alpha^3}$ contributions. In addition to this,
it models the emission of an arbitrary number of additional photons using the QED Parton Shower algorithm.

\begin{figure}[hb]
\center
\includegraphics[width=1.1\columnwidth]{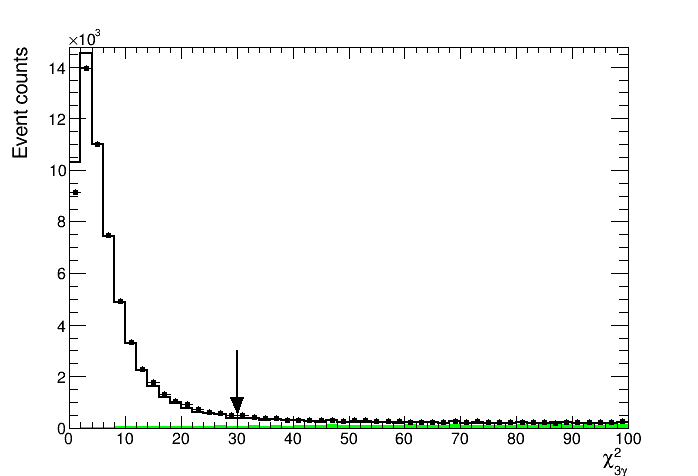}
\caption{\label{Chi2}
The distribution of $\chi^2_{3\gamma}$ at $E = 1020.055$ MeV for the data (dots with error bars)
and simulation of the signal and the background processes (open histogram).
The shaded histogram represents the contribution of the multiphoton hadronic background.
The simulated distributions are normalized to the number of expected events.
The arrow indicates the boundary of the selection condition.}
\end{figure}

The main hadronic background, which also gives a three-photon final state, is the $e^+e^-\to\pi^0\gamma$ process. 
In addition to it, we also study the contribution to the background from the hadronic processes with multiphoton final state
$e^+e^- \to K_{S}K_{L}$, $e^+e^- \to 2\pi^0\gamma$, $e^+e^-\to\eta\gamma\to3\pi^0\gamma$. The cosmic ray background
is studied by analysing the event time measured in the electromagnetic calorimeter and found to be negligible.

Figure~\ref{Chi2} shows the $\chi^2_{3\gamma}$ distribution for data and MC simulation 
at $E = 1020.055$ MeV after applying the selection conditions. 
All distributions for the MC simulation are normalized to the expected number of events.
The shares of their contributions to the signal region $\chi^2_{3\gamma} < 30$ is following:
$e^+e^-\to\eta\gamma\to3\gamma$ - 60.0\%, QED processes - 26.1\%, $e^+e^-\to\pi^0\gamma$ - 12.4\%, multiphoton hadronic processes - 1.4\%.
Since the $e^+e^- \to \gamma \gamma (\gamma)$ and $e^+e^-\to\pi^0\gamma$,
as well as the studied process $e^+e^-\to\eta\gamma\to3\gamma$, have three photons in the final state, their
$\chi^2_{3\gamma}$ distributions have a peak at low values of $\chi^2_{3\gamma}$. For this reason, these processes
give the dominant contribution to the background in the signal region.

\section{Fitting the $M_{\gamma\gamma}$ distribution}

\begin{figure*}[ht]
\center
\includegraphics[width=0.32\textwidth]{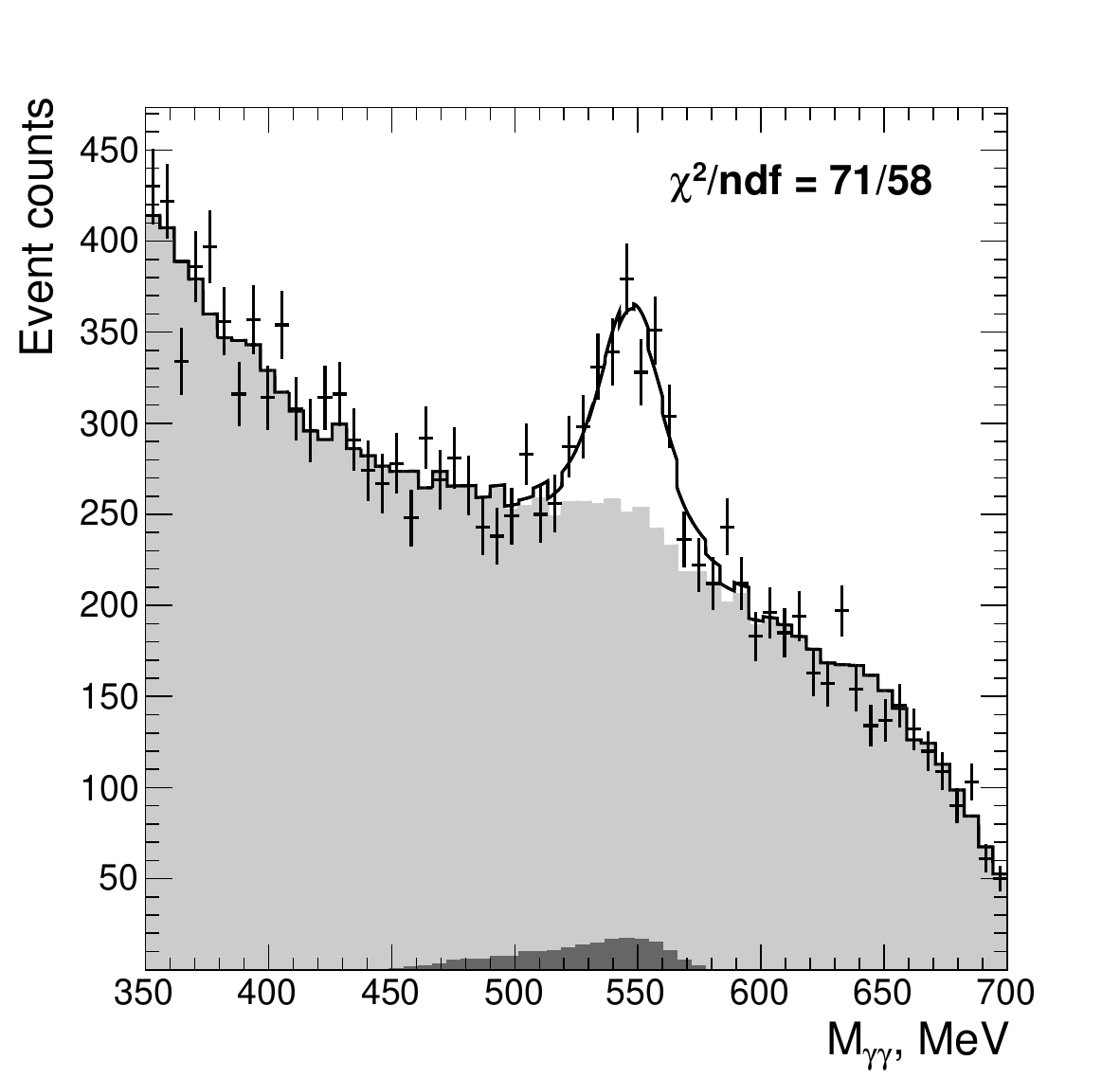}\hfill 
\includegraphics[width=0.32\textwidth]{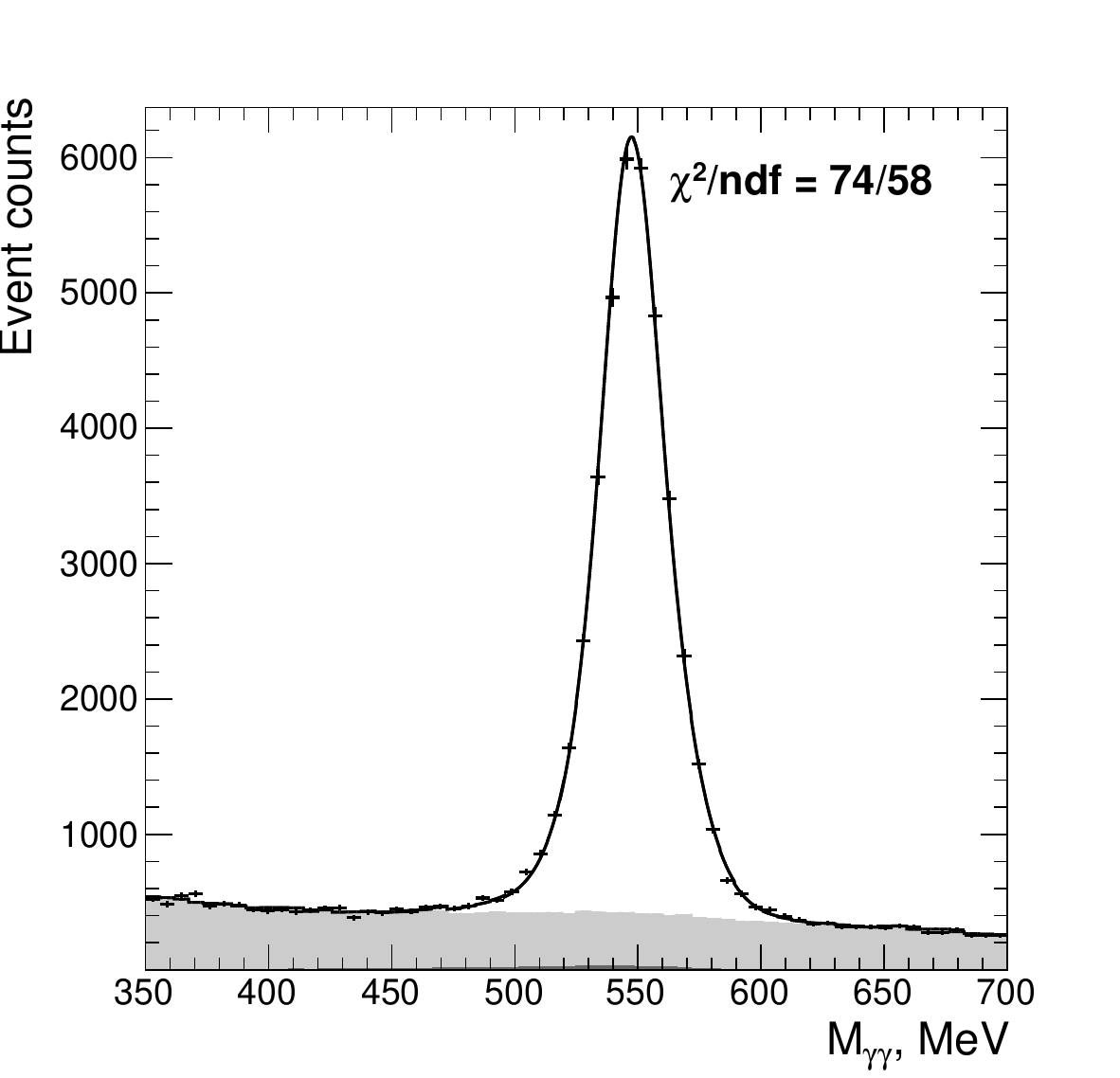}\hfill 
\includegraphics[width=0.32\textwidth]{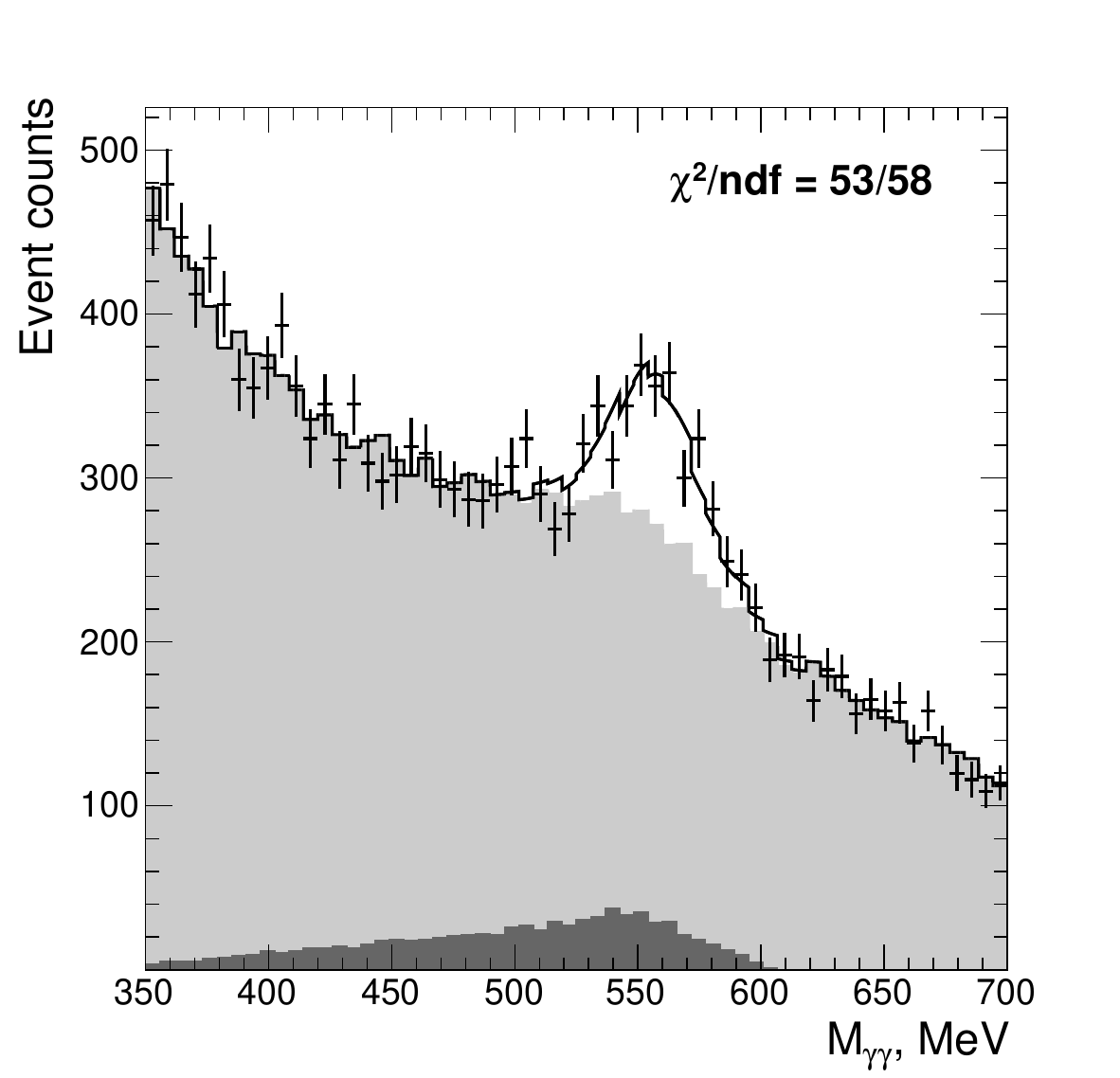}\\
\caption{\label{Mgg}
The $M_{\gamma\gamma}$ distribution for data events (points with error bars) at $E = 1000.039$ MeV
(left), $E = 1020.055$ MeV (middle), $E = 1049.992$ MeV (right). The solid curve represents the result 
of the fit described in the text. The background distribution is shown as stacked histograms for 
$F_{bkg}^{I}$ (in light gray) and $F_{bkg}^{II}$ (in dark gray).
}
\end{figure*}

The $M_{\gamma\gamma}$ distributions for the energy points corresponding to the maximum of the $\phi(1020)$ resonance and its
left and right tails are shown in Fig.~\ref{Mgg}.
To determine the number of signal events ($N_{sig}$) at each energy point, we perform a binned likelihood 
fit to the $M_{\gamma\gamma}$ distribution with a sum of signal ($F_{sig}$) and background ($F_{bkg}$) distributions. 
The background distribution is composed of two components $F_{bkg}^{I}$ and $F_{bkg}^{II}$.
The distributions $F_{bkg}^{I}$ and $F_{bkg}^{II}$ are shown in Fig.~\ref{Mgg} as stacked histograms.
They content background events with $E_3 < m_{\eta}^{2}/E$ and $E_3 > m_{\eta}^{2}/E$, respectively.

The signal distribution is obtained by fitting the mass spectra for the simulated signal events
by a sum of three Gaussian functions.
When fitting the data, the $F_{sig}$ function obtained from the simulation is modified 
to account for possible difference between the data and the simulation in the detector response.
A common mass shift $\Delta M$ and a smearing $\Delta \sigma^2$ 
are added to each Gaussian mean and variance, respectively, in the following way:
$M_{\gamma\gamma}^{data} = M_{\gamma\gamma}^{MC} + \Delta M$ and $\sigma^2_{data} = \sigma^2_{MC} + \Delta \sigma^2$.
The values of $\Delta M$ and $\Delta \sigma^2$ are determined by fitting the data
$M_{\gamma\gamma}$ distributions in the energy region $E = 1017 - 1023$ MeV. They are found to be
$\Delta M = -0.77 \pm 0.05 (-0.42 \pm 0.08)$ MeV and $\Delta \sigma^2 = 1.11 \pm 1.18 (1.21 \pm 1.88)$ MeV$^{2}$ 
for the 2024 (2018) energy scan. The parameters $\Delta M$ and $\Delta \sigma^2$ are fixed at these values. The variation of these
parameters within the errors is used to estimate the associated systematic uncertainty.

The background distributions $F_{bkg}^{I}$ and $F_{bkg}^{II}$ are described by histograms obtained from 
simulation of background processes $e^+e^- \to \gamma \gamma (\gamma)$, $e^+e^-\to\pi^0\gamma$,
$e^+e^- \to K_{S}K_{L}$, $e^+e^- \to 2\pi^0\gamma$ and $e^+e^-\to\eta\to3\pi^0\gamma$.
To model possible inaccuracies of such a description, we 
introduce two additional parameters $\alpha$ and $\beta$ and
parametrize the background distribution $F_{bkg}$ as follows
\begin{equation}
F_{bkg} = \alpha\left[\beta \cdot F_{bkg}^{I} + ((1-\beta) \cdot D + 1) \cdot F_{bkg}^{II}\right],
\end{equation}
where $D$ is the ratio of numbers of expected events in $F_{bkg}^{I}$ and $F_{bkg}^{II}$.
The parameters $\alpha$ and $\beta$ model the possible inaccuracy of MC predictions for 
the number of background events and parameter $D$, respectively.
The values of parameters $\alpha = 1$ and $\beta = 1$ correspond to the expected background distribution.

Usually, the photon with the misreconstructed energy has the lowest energy out of the three photons
selected for the kinematic fitting.
Since we choose a pair of photons for $M_{\gamma\gamma}$ depending on the energy of this photon $E_3$,
we cannot fully rely on simulation to determine the parameter $D$. Instead, the parameter
$\beta$ is determined by fitting the data $M_{\gamma\gamma}$ distributions in the energy region
out of the $\phi(1020)$ resonance peak $E = 980 - 1013$ MeV and $E = 1025 - 1060$ MeV. For all
energy points from this region, the obtained values for the parameter $\beta$ are consistent with unity.
When we determine $N_{sig}$, the parameter $\beta$ is fixed at its average value  $\beta = 1.000 \pm 0.004$
and its error is used to estimate the associated systematic uncertainty.
The parameter $\alpha$ is set to be free. The obtained values for it have no visible energy dependence.
It average value is $\alpha = 0.984\pm0.003$.

The results of the fit for three energy points are shown in Fig.~\ref{Mgg}. 
The fit quality is reasonably good in the whole studied c.m. energy
range. The procedure of choosing a pair of photons for $M_{\gamma\gamma}$ notably distorts the background 
distribution mostly because the $F_{bkg}^{II}$ distribution looks like a bump at the $\eta$ meson mass~(Fig.~\ref{Mgg}).
The possible inaccuracy of modeling this distortion is parametrized by the parameter $\beta$ and gives
the dominant contribution to the systematic uncertainty in $N_{sig}$. 

The obtained numbers of signal events for each energy point are listed in Table~\ref{tab2}
with statistical and systematic errors. The systematic error is obtained by variation of the parameters $\Delta M$, $\Delta \sigma^2$ 
and $\beta$ within their errors.

\section{Detection efficiency}
\begin{figure}[t]
\center
\includegraphics[width=1.1\columnwidth]{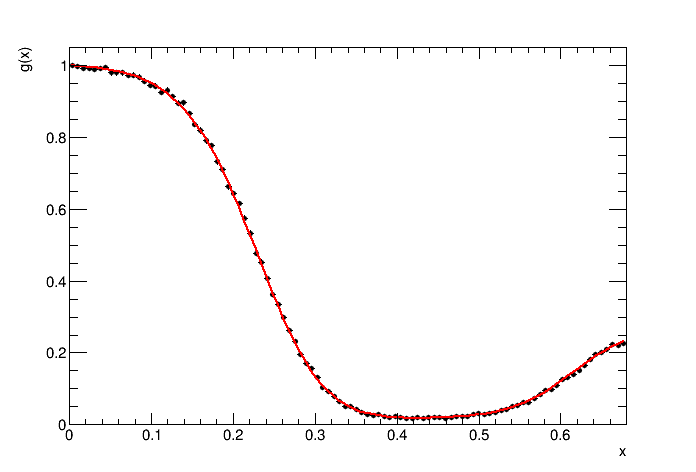}
\caption{\label{g_x}
The $x$ dependence of the detection efficiency obtained from simulation is shown by 
points with error bars. The solid curve is the result of approximation by a smooth function.
}
\end{figure}

The detection efficiency of $e^+e^-\to\eta\gamma\to3\gamma$ events
is determined using MC simulation as a function of $E$ and $x = 2 E_{r}/E$, where $E_{r}$
is the energy of the extra photon emitted from the initial state.
It is parametrized as $\varepsilon_{r}(E,x) = \varepsilon_{0}(E)g(E,x)$, where $\varepsilon_{0}(E) = \varepsilon_{r}(E,0)$ and
$g(E,0) = 1$. Since the studied energy region is narrow, the dependence of $g$ on $E$ can be neglected.
In this case, all variations of the experimental conditions 
(dead calorimeter channels, beam background, etc.) are accounted for in $\varepsilon_{0}(E)$.
The dependence of $g$ on $x$ is obtained using MC simulation and then 
approximated by a smooth function. The result of the approximation is shown in Fig.~\ref{g_x}.
The decrease of the efficiency at $x=0.1-0.3$ is due to the requirement of energy and momentum 
balance in an event (the condition $\chi^2_{3\gamma} < 30$).
The increase near the maximum allowed value of $x$ 
corresponds to the radiative return to the threshold of the reaction $e^+e^-\to\eta\gamma$ when
the photon recoiling against the $\eta$ meson becomes soft.
The initial state radiation photon together with two photons from the $\eta$ decay provide a good
energy-momentum balance, and such an event passes the selection conditions.
In spite of this,  the number of events from this kinematic region is negligible 
due to the low probability of emitting a hard photon and the very small $e^+e^-\to\eta\gamma$
cross section near the threshold.

The imperfect MC simulation of the detector response for photons leads
to an inaccuracy in the detection efficiency determined using the MC simulation. 
To evaluate the efficiency corrections, we vary the boundaries of the selection conditions.
The correction is defined as a double ratio $1+\delta_{cut} = (N^{data}/N_{MC}) / (N^{data}_{0}/N^{MC}_{0})$, where
$N^{data}$ ($N^{data}_{0}$) and $N^{MC}$ ($N^{MC}_{0}$) are 
the number of signal events in the data and in the signal MC simulation, respectively, for the modified (baseline)
selection condition. With the correction defined in this way, the corrected efficiency $\varepsilon_0$
is related to the efficiency obtained from the MC simulation $\varepsilon_0^{MC}$ as follows
$\varepsilon_0 = \varepsilon_0^{MC} / (1+\delta_{cut})$.

To estimate the systematic uncertainty associated with the condition on the polar angle
of photons accepted for the kinematic fit ($\theta_{0} < \theta_{\gamma} < 180^{\circ} - \theta_{0}$
with $\theta_{0} = 27^{\circ}$), we change $\theta_{0}$ to $36^{\circ}$ and $45^{\circ}$. 
The step of the $\theta_{0}$ variation corresponds to the angular size of the calorimeter crystal. 
The signal efficiency decreases by 28 (50)\% for $\theta_{0} = 36^{\circ} (45^{\circ})$, 
while the maximum deviation of the correction obtained for the different $\theta_{0}$ 
does not exceed 0.6\%. This deviation is taken as an estimate of the systematic uncertainty 
due to the condition on the photon polar angle.

The fraction of signal events rejected by the condition $\chi^2_{3\gamma} < 30$ is 7.7\% at the maximum of the $\phi(1020)$ resonance.
When the selection condition is relaxed to $\chi^2_{3\gamma} < 100$~$(800)$, the fraction of rejected events decreases to 3.0\% (0.1\%).
To evaluate the efficiency correction, we relax the condition from $\chi^2_{3\gamma}<30$ 
to $\chi^2_{3\gamma}<800$ and impose additional constraints requiring
exactly three photons in an event and no signal in the muon system.
In a signal event with $\chi^2_{3\gamma} > 100$, typically
at least one photon hits the dead counter of the electromagnetic calorimeter. 
This leads to a misreconstruction of the photon energy and, as a consequence,
to an incorrect reconstruction of the kinematics of the entire event. 
For this reason, the peak at the $\eta $ meson mass in the $M_{\gamma\gamma}$ distribution 
becomes very broad, making it impossible to separate the signal events from the background.
At the same time, the angles of such photons are measured with sufficient accuracy.
To improve the resolution of the reconstructed $\eta$ meson mass for events with 
$\chi^2_{3\gamma} > 100$, we perform a special kinematic fit using 
only the measured photon angles and ignoring their measured energies.
The variation of the correction when changing the condition on $\chi^2_{3\gamma}$ 
from 100 to 800 is taken as a systematic uncertainty.
The efficiency correction is determined at each energy point in the energy range $E=1017-1023$ MeV.
For the 2024 energy scan $\delta_{\chi^2}$ does not depend on energy point and
its value is $\delta_{\chi^2} = 1.2 \pm 0.3$ \%. 
For the 2018 energy scan $\delta_{\chi^2}$ varies from $-1.1$\% to  $3.0$\% 
with systematic uncertainty estimated to be 0.6\%.

A photon converted in the material before 
the tracking system is reconstructed as a charged track. Since events with charged tracks
are rejected by our selection conditions, the data-MC simulation difference in 
the conversion probability leads to a systematic shift in the measured cross section.
This difference was studied in Ref.~\cite{SNDkskl}, where the conversion probability
was found to be greater than in the simulation by ($0.23 \pm 0.02$)\%$/$sin$\theta_{\gamma}$,
where $\theta_{\gamma}$ is the photon polar angle. The efficiency correction
for the process $e^+e^-\to\gamma\gamma$ is calculated to be $(0.53 \pm 0.04)$\% and was applied 
to the measured luminosity. The correction for the process $e^+e^-\to\eta\gamma$
is calculated to be $\delta_{conv} = (0.86 \pm 0.06)$\%.

The corrected detection efficiencies $\varepsilon_0$ are listed in Table~\ref{tab2}.

\section{Fitting the measured cross section}

\begin{table}[b]
\caption{\label{FitTab} Results of the fit described in the text.}
\begin{ruledtabular}
\begin{tabular}{ccccc}
Parameter & Solution 1 & Solution 2 & Solution 3 & Solution 4\\
\hline
$\sigma_{\phi}$, nb & $60.69^{+0.36}_{-0.39}$ & $56.39^{+0.35}_{-0.31}$ & $57.24^{+0.42}_{-0.50}$ & $61.19^{+0.34}_{-0.38}$\\
$\varphi_{\phi}$, deg. & $192 \pm 6$ & $124 \pm 5$ & $150 \pm 8$ & $215 \pm 5$\\
$\sigma_{\rho^{\prime}}$, pb & $3.6^{+3.0}_{-2.2}$ & $1.5^{+2.2}_{-1.5}$ & $28.5^{+6.6}_{-6.8}$ & $32.7^{+8.0}_{-8.2}$\\ 
$\sigma_{\phi^{\prime}}$, pb & $5.4^{+8.7}_{-3.5}$ & $4.6^{+6.2}_{-3.0}$ & $3.5^{+2.7}_{-3.5}$ & $3.0^{+4.5}_{-3.0}$\\ 
$\varphi_{\rho^{\prime}}$, deg. & $28 \pm 31$ & $39 \pm 28$ & $-103 \pm 9$ & $-147 \pm 8$\\
$\Delta E_{18}$, keV & $40 \pm 10$ & $41 \pm 10$ & $40 \pm 10$ & $40 \pm 10$ \\
$\Delta E_{24}$, keV & $-213 \pm 15 $ & $-208 \pm 15$ & $-211 \pm 14$ & $-214 \pm 15$\\
$\chi^2/ndf$  & 35.2/37 & 37.4/37 & 37.6/37 & 36.5/37\\ 
\end{tabular}
\end{ruledtabular}
\end{table}

The visible cross section of the process $e^+e^-\to\eta\gamma$ is determined as
\begin{equation}
\label{viscrs0}
\sigma_{vis}(E) = \frac{N_{sig}}{IL \cdot B(\eta\to 2\gamma)},
\end{equation}
where IL is the integrated luminosity and $B(\eta\to 2\gamma) = 39.36\pm0.18$~\%~\cite{pdg}.
It is related to the Born cross section $\sigma(E)$ as follows:
\begin{equation}
\label{viscrs}
\sigma_{vis}(E) = \int\limits_{0}^{x_{max}} \varepsilon_{0}(E)g(E,x) F(x,E) \sigma(E\sqrt{1-x})dx~,
\end{equation}
where $F(x,E)$ is a so-called radiator function~\cite{radcor} describing the probability 
to emit an extra photon with the total energy $xE/2$ from the initial state,
$x_{max} = 1 - m_{\eta}^2/E^2$. The expression (\ref{viscrs}) can be rewritten
in the conventional form:
\begin{equation}
\label{viscrs1}
\sigma_{vis}(E) = \varepsilon_{0}(E)\,\sigma(E)\,(1+\delta(E))~,
\end{equation}

To take into account the beam energy spread, it is necessary to perform a convolution 
of the visible cross section (\ref{viscrs1}) with a Gaussian function describing 
the c.m. energy distribution of the electron-positron pair. Since the energy spread is 
much smaller than the width of the $\phi(1020)$ resonance, we use an approximate 
expression for the convolution

\begin{eqnarray}
\sigma_{ vis}(E) &\Longrightarrow&
\sigma_{ vis}(E)+ 
\frac{1}{2}\frac{d^2\sigma_{ vis}}{dE^2}(E)\sigma_E^2\nonumber \\ 
&=&\sigma_{\rm vis}(E)(1+\delta_E(E)) 
\label{conv} 
\end{eqnarray}

The uncertainty in the collider energy listed in Table~\ref{tab1} effectively increases the uncertainty 
in the measured visible cross section. When fitting the cross section energy dependence, 
the following term is added in quadrature to the statistical error of
$\sigma_{vis}$:  
\begin{equation} 
\delta E \frac{d\sigma_{vis}}{dE}(E), 
\label{desys} 
\end{equation} 
where $\delta E$ is the uncertainty of the energy measurement of the corresponding energy point. 
In the energy range $E=1016-1023$ MeV, this additional uncertainty is comparable to the statistical error.

Energy dependence of the $e^+e^-\to\eta\gamma$ Born cross section
is described using the vector meson dominance model, which, 
in addition to the dominant amplitude of the $\phi$ meson, includes 
the amplitudes of the $\rho$ and $\omega$ mesons as well as amplitudes
describing the contribution of the high-lying vector resonances
$\rho(1450)$ ($\rho^{\prime}$) and $\phi(1680)$ ($\phi^{\prime}$): 

\begin{eqnarray}
\label{crs1}
\sigma_{\eta\gamma}(E) &=& \frac{q(E)^3}{E^3} \left| 
\sum\limits_{V=\rho,\omega,\phi,\rho^\prime,\phi^\prime} A_{V}(E)\right|^2,\\
A_V(E) &=& \frac{m_V \Gamma_V e^{i\varphi_V}}{D_V(E)} \sqrt{ \frac{m_V^3}
{q(m_V)^3}\sigma_V}~, \\
D_V(E) &=& m^2_V -E^2 - i E\Gamma_V(E), \\
q(E) &=& \frac{E}{2} \left( 1 - \frac{m^2_{\eta}}{E^2} \right)~,
\end{eqnarray}
where $m_V$ is the resonance mass, $\Gamma_V(E)$ is its energy-dependent total width 
($\Gamma_V\equiv \Gamma_V(m_V)$), $\sigma_V$ is the Born cross section of the process
$e^+e^- \to V \to \eta\gamma$ at $E = m_V$,
$\varphi_V$ are interference phases relative to the $\rho$ meson ($\varphi_{\rho} \equiv 0$). 

The quark model predicts $\varphi_{\omega} = 0^{\circ}$ and $\varphi_{\phi} = 180^{\circ}$~\cite{achasov},
which agrees well with measurements~\cite{SNDetg2007}. Assuming that the resonances $\rho(1450)$ and $\phi(1680)$ 
belong to the same $2^3S_1$ nonet and taking into account the lattice 
QCD predictions of their almost ideal mixing~\cite{dudek}, we can expect the same pattern for their relative phases. 
For these reasons, the phase between the $\rho$ and $\omega$ mesons is fixed at $\varphi_{\omega} = 0^{\circ}$,
and $\varphi_{\phi^{\prime}} = \varphi_{\rho^{\prime}} + 180^{\circ}$.

To determine the contribution of high-lying vector resonances, the cross section
measured by SND~\cite{SNDetg23} in the energy range $E = 1070 - 2000$ MeV
is added to the fit. The masses and widths of all vector resonances are fixed at
their PDG values~\cite{pdg}. The parameters $\sigma_{\rho}$ and $\sigma_{\omega}$ are fixed
at the values  taken from the SND measurement~\cite{SNDetg2007},
\begin{eqnarray}
\sigma_{\rho} &=& (0.322 \pm 0.0034 \pm 0.019)~\text{nb},\\
\sigma_{\omega} &=& (0.744 \pm 0.075 \pm 0.027)~\text{nb}.
\end{eqnarray}
The free fit parameters are $\sigma_{\phi}$, $\varphi_{\phi}$, $\sigma_{\rho^{\prime}}$, $\varphi_{\rho^{\prime}}$,
$\sigma_{\phi^{\prime}}$ and the shifts of the energy scale for the 
2024 scan $\Delta E_{24}$ and for the 2018 scan $\Delta E_{18}$. As was discussed above, the parameter $\Delta E_{18}$ is
constrained by a Gaussian penalty term corresponding to $\Delta m^{FSI}_{\phi} = 0.035 \pm 0.012$ MeV.
To evaluate the systematic uncertainty associated with the energy scale, the constraint in penalty 
term is changed to $\Delta m_{\phi} = 0.020 \pm 0.012$ MeV.

\begin{table*}[th]
\caption{The center-of-mass energy ($E$), detection efficiency
($\varepsilon_0$), number of signal events ($N_{sig}$), radiative
correction ($1+\delta$), correction for the energy spread
($1+\delta_E$), Born cross section for the process $e^+e^-\to \eta\gamma$
($\sigma$). The first error in the number of events and cross section is
statistical, the second is systematic.
\label{tab2}}
\begin{ruledtabular}
\begin{tabular}{cccccc}
$E$, GeV & $\varepsilon_0$, \% & $N_{sig}$ & $1+\delta$ & $1+\delta_E$ & $\sigma$, nb\\
\hline   
\multicolumn{6}{c}{2024 energy scan} \\
980.189  & 53.4 & $155    \pm  42    \pm 23$ & 0.881 & 1.000 & $0.41  \pm   0.11 \pm 0.06$ \\   
984.124  & 53.1 & $238    \pm  44    \pm 24$ & 0.872 & 1.000 & $0.59  \pm   0.11 \pm 0.06$ \\   
990.032  & 53.0 & $161    \pm  41    \pm 17$ & 0.856 & 1.000 & $0.47  \pm   0.12 \pm 0.05$ \\   
1000.039 & 52.7 & $712    \pm  62    \pm 33$ & 0.823 & 1.001 & $1.03  \pm   0.09 \pm 0.05$ \\   
1002.038 & 53.3 & $537    \pm  47    \pm 16$ & 0.815 & 1.001 & $1.58  \pm   0.14 \pm 0.05$ \\   
1006.034 & 53.3 & $769    \pm  51    \pm 17$ & 0.797 & 1.002 & $2.00  \pm   0.13 \pm 0.05$ \\   
1010.091 & 53.6 & $1376   \pm  57    \pm 15$ & 0.774 & 1.006 & $3.88  \pm   0.16 \pm 0.06$ \\   
1013.099 & 53.0 & $2377   \pm  67    \pm 16$ & 0.752 & 1.011 & $7.19  \pm   0.21 \pm 0.09$ \\   
1016.110 & 52.6 & $10372  \pm  121   \pm 27$ & 0.726 & 1.022 & $21.68 \pm   0.29 \pm 0.27$ \\   
1017.073 & 52.9 & $16651  \pm  149   \pm 32$ & 0.719 & 1.018 & $32.55 \pm   0.37 \pm 0.37$ \\   
1018.111 & 53.3 & $26770  \pm  183   \pm 36$ & 0.719 & 0.988 & $50.71 \pm   0.45 \pm 0.53$ \\   
1019.073 & 52.6 & $41526  \pm  225   \pm 50$ & 0.740 & 0.960 & $59.79 \pm   0.44 \pm 0.75$ \\   
1020.055 & 52.9 & $36168  \pm  212   \pm 46$ & 0.790 & 0.990 & $46.89 \pm   0.40 \pm 0.49$ \\   
1021.005 & 52.8 & $19774  \pm  159   \pm 31$ & 0.863 & 1.015 & $30.13 \pm   0.31 \pm 0.32$ \\   
1023.079 & 51.7 & $8694   \pm  112   \pm 22$ & 1.055 & 1.016 & $11.60 \pm   0.17 \pm 0.13$ \\   
1025.003 & 51.9 & $2925   \pm  70    \pm 12$ & 1.257 & 1.006 & $5.59  \pm   0.14 \pm 0.06$ \\   
1025.058 & 52.0 & $3354   \pm  74    \pm 12$ & 1.263 & 1.008 & $5.81  \pm   0.14 \pm 0.06$ \\   
1030.072 & 52.5 & $1568   \pm  61    \pm 12$ & 1.871 & 1.003 & $1.570 \pm   0.069\pm 0.020$\\   
1040.047 & 52.7 & $518    \pm  47    \pm 8 $ & 3.477 & 1.001 & $0.343 \pm   0.036\pm 0.006$\\   
1049.992 & 52.4 & $713    \pm  69    \pm 15$ & 5.672 & 1.000 & $0.124 \pm   0.017\pm 0.004$\\   
1060.206 & 53.2 & $84     \pm  25    \pm 2$  & 8.505 & 1.000 & $0.079 \pm   0.025\pm 0.004$\\   
\multicolumn{6}{c}{2018 energy scan} \\                                                         
1005.999 & 52.9 & $598   \pm  44  \pm 10$  & 0.797 & 1.002 & $2.14   \pm   0.16  \pm 0.04$ \\   
1016.843 & 52.5 & $6482  \pm  95  \pm 15$  & 0.722 & 1.014 & $25.97  \pm   0.48  \pm 0.33$ \\   
1017.959 & 51.2 & $7711  \pm  100 \pm 12$  & 0.718 & 1.003 & $42.30  \pm   0.72  \pm 0.49$ \\   
1019.101 & 52.6 & $20990 \pm  161 \pm 27$  & 0.732 & 0.972 & $58.02  \pm   0.55  \pm 0.74$ \\   
1019.982 & 52.8 & $21546 \pm  163 \pm 26$  & 0.770 & 0.982 & $51.91  \pm   0.55  \pm 0.67$ \\   
1020.932 & 52.4 & $8747  \pm  105 \pm 12$  & 0.836 & 1.008 & $35.36  \pm   0.80  \pm 0.41$ \\   
1022.114 & 51.7 & $4744  \pm  80  \pm 9 $  & 0.938 & 1.012 & $19.96  \pm   0.43  \pm 0.23$ \\   
1022.929 & 53.0 & $2343  \pm  58  \pm 5 $  & 1.015 & 1.013 & $13.31  \pm   0.40  \pm 0.16$ \\   
\end{tabular}
\end{ruledtabular}     
\end{table*}

\begin{figure*}[h]
\center
\includegraphics[width=0.5\textwidth]{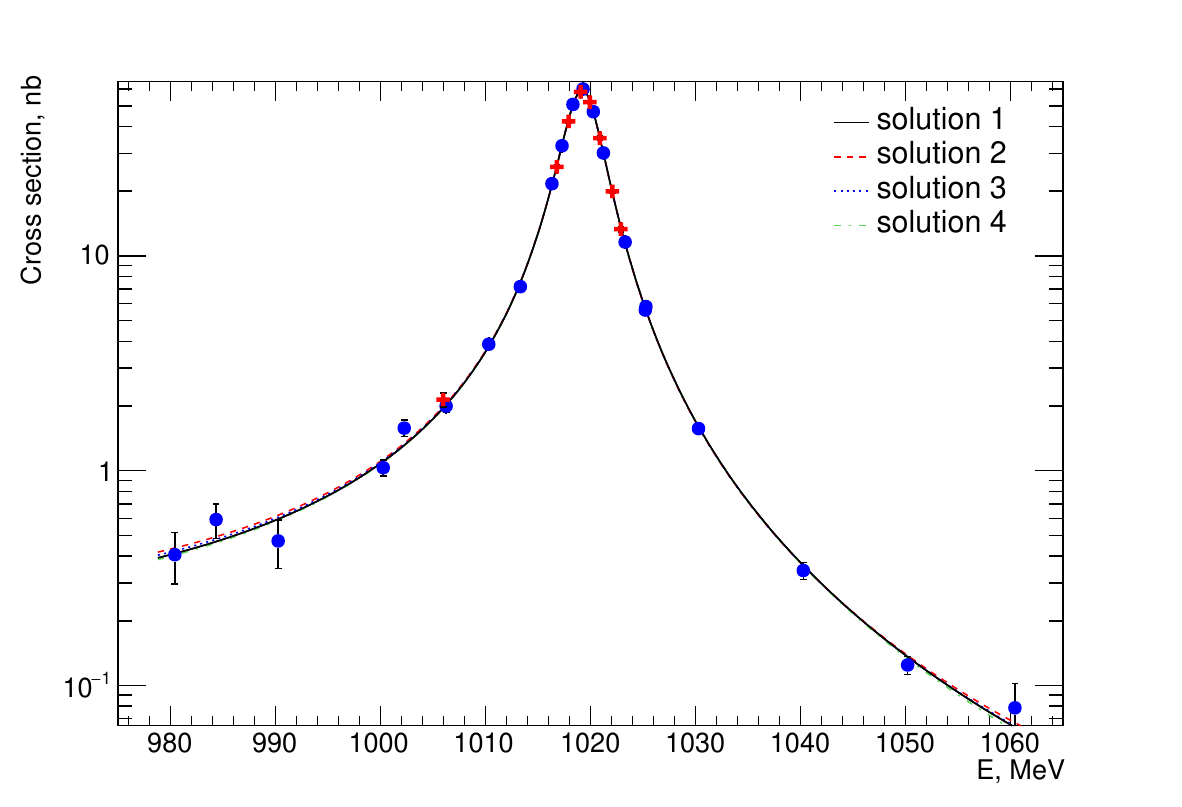}\hfill 
\includegraphics[width=0.5\textwidth]{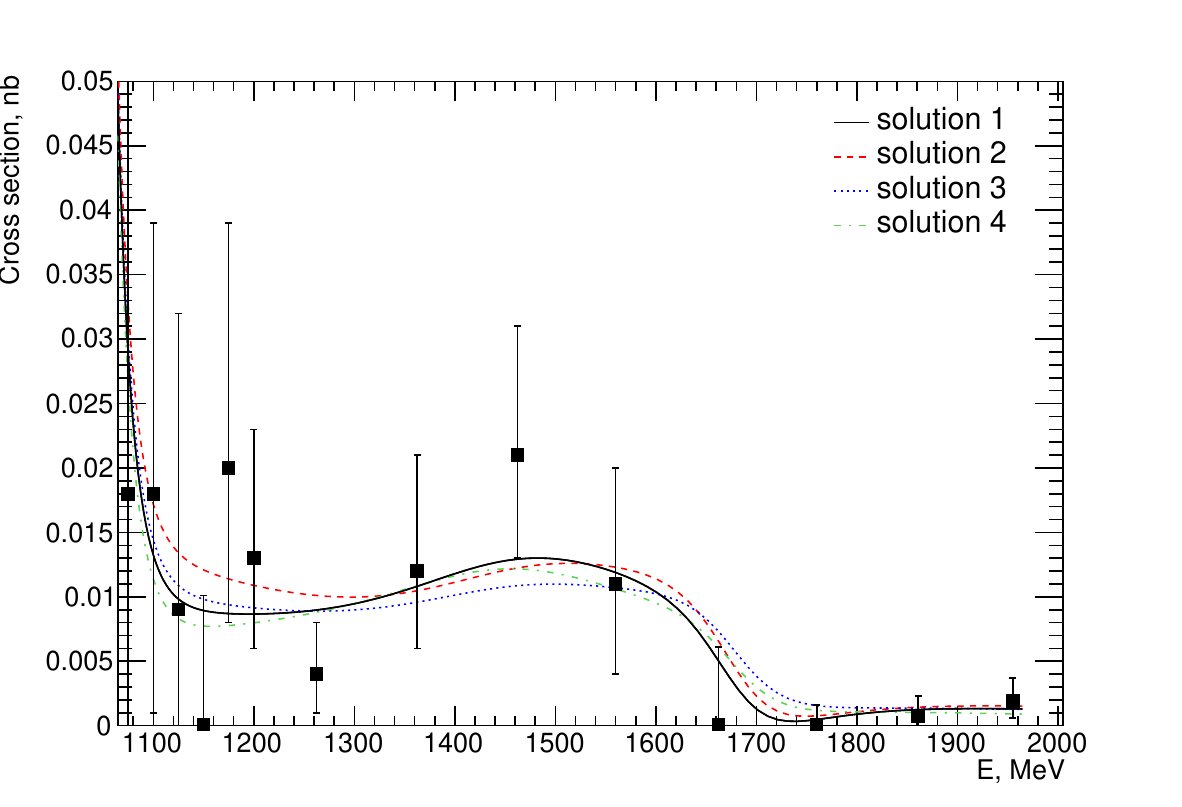}\hfill 
\caption{\label{crs_fit}
The Born cross section for the $e^+e^- \to \eta\gamma$ process. Left panel shows
the cross section measured in this work for the  2024 (blue dots) and 2018 (red crosses)
energy scans. Right panel shows the cross section measured by SND in Ref.~\cite{SNDetg23}.
The black solid, red dashed, blue dotted and green dash-dotted curves represent the result of the fit described in the text
for solution 1, 2, 3, and 4, respectively.
The c.m. energy is corrected by the value of the energy scale shift (see Table~\ref{FitTab}).
}
\end{figure*}

\begin{figure*}[ht]
\center
\includegraphics[width=0.5\textwidth]{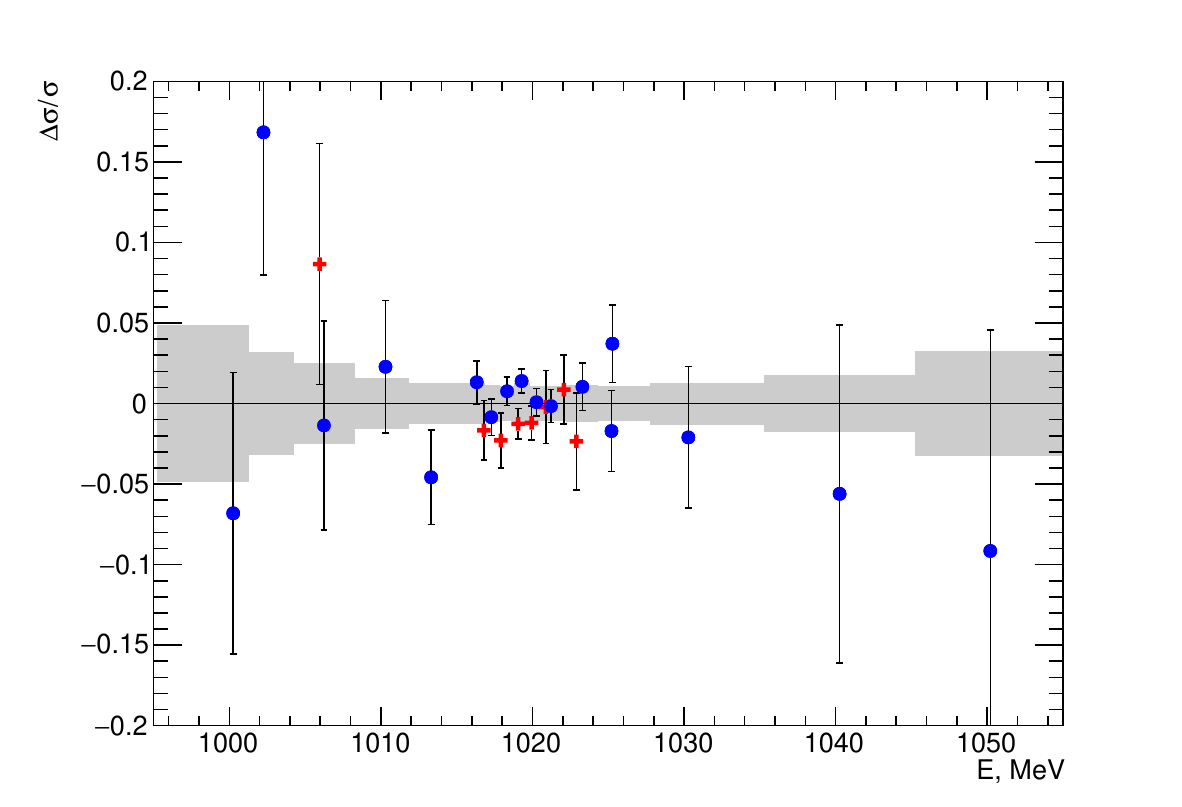}\hfill 
\includegraphics[width=0.5\textwidth]{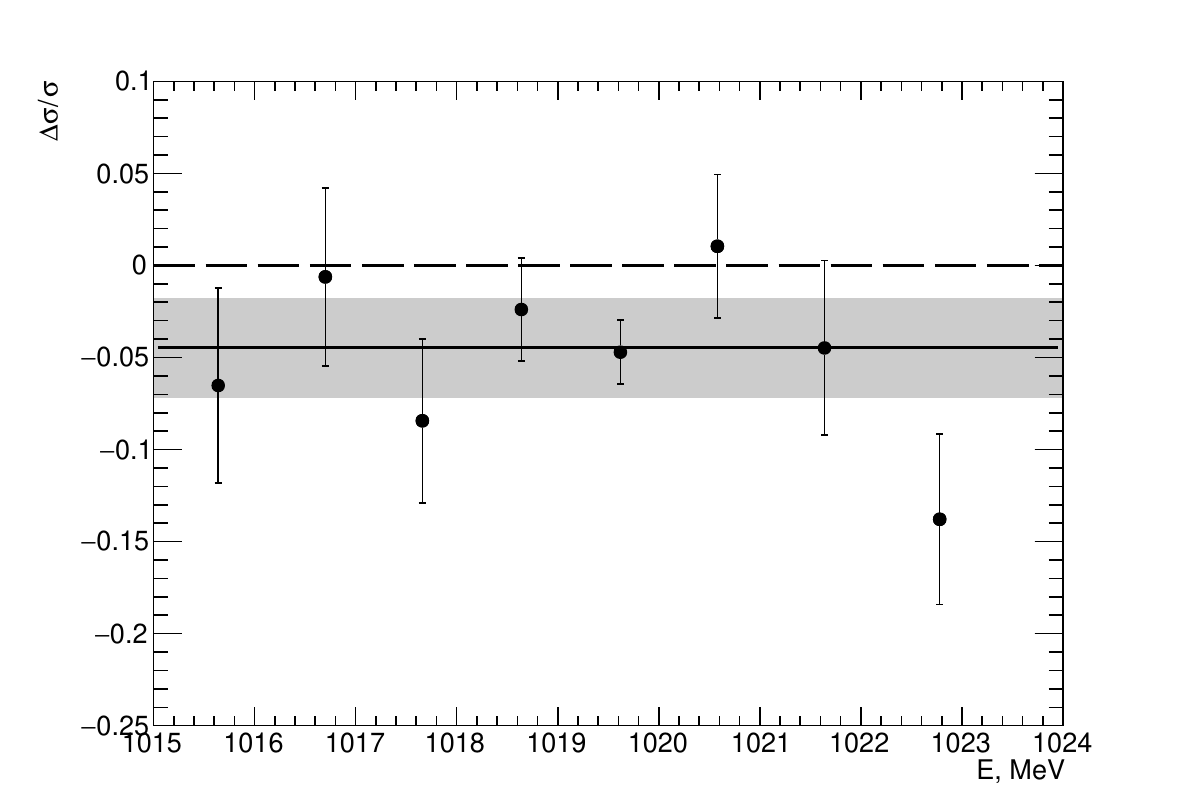}\hfill \\
\includegraphics[width=0.5\textwidth]{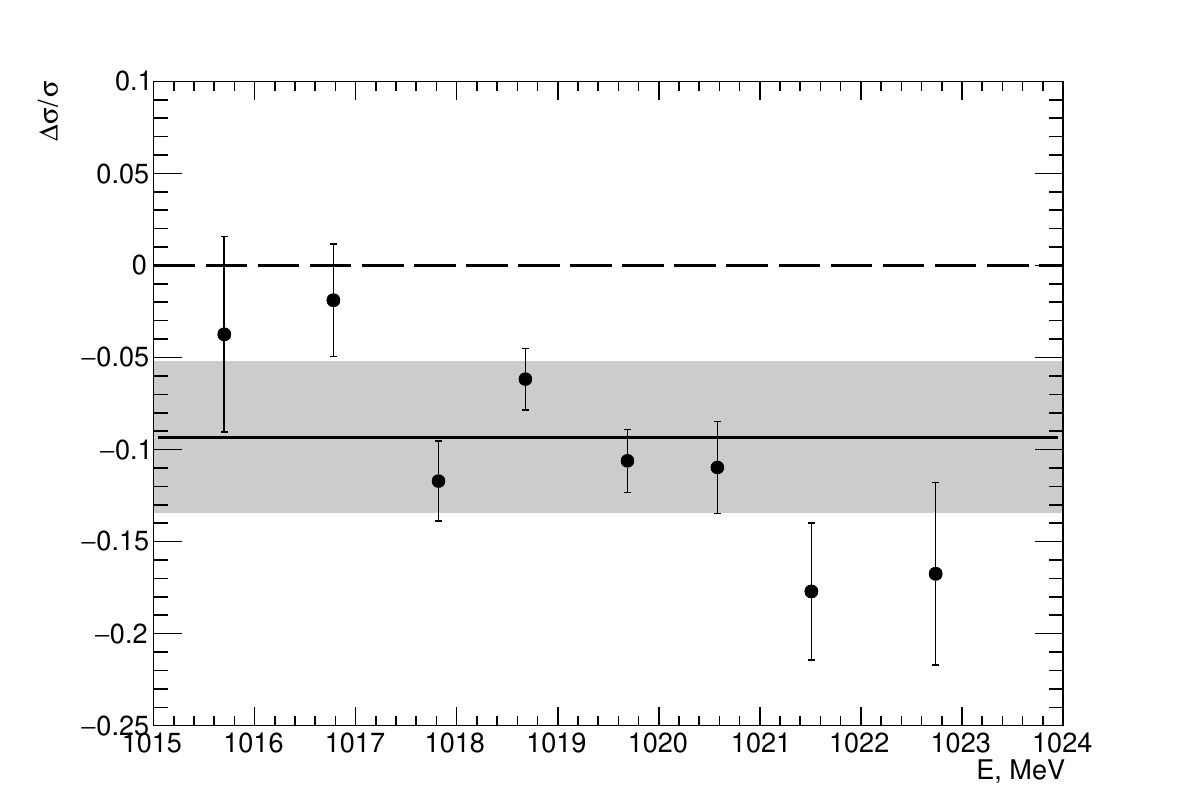}\hfill 
\includegraphics[width=0.5\textwidth]{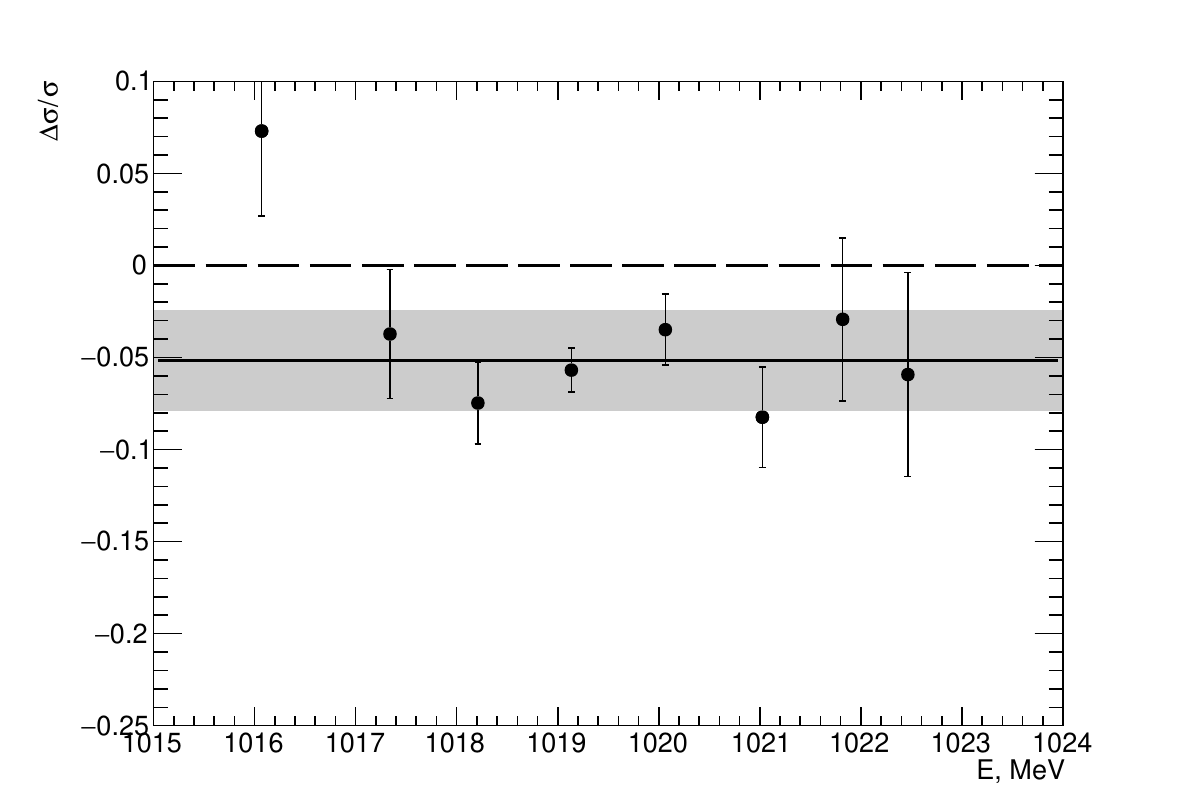}\hfill 
\caption{\label{crs_diff}
The relative differences between the measured 
$e^+e^- \to \eta\gamma$ Born cross section and the result of the fit described in the text for
this work (top left), for the SND measurement using the decay channel $\eta\to3\pi^0$ in Ref.~\cite{SNDetg2007} (top right),
for the CMD-2 measurement in Ref.~\cite{CMDetg2001} (bottom left) and for the SND measurement in Ref.~\cite{SNDetg2000} (bottom right).
The solid line represents the average
value of the relative difference. The shaded region represents the systematic
uncertainty of corresponding measurement. The c.m. energy for the measurement from this work 
is corrected by the value of the energy scale shift (see Table~\ref{FitTab}).
}
\end{figure*}

The result of the fit is presented in Table~\ref{FitTab} and shown in Fig.~\ref{crs_fit}. 
We found four different solutions with very close and good fit quality $\chi^2/ndf$ = (35-38)/37,
where $ndf$ is the number of degrees of freedom.
This is a consequence of 
the well-known problem that the interfering Breit-Wigner resonances have ambiguities in 
the determination of their parameters~\cite{malyshev}. As discussed above, 
the quark model predicts the relative phase $\varphi_{\phi} = 180^{\circ}$. 
The only first solution presented in Table~\ref{FitTab} has relative phase $\varphi_{\phi}$
compatible with this value.
For this reason, we choose it as a nominal one.
We use this solution for all calculations with the theoretical model, unless otherwise explicitly stated.
It is also worth noting that in both experimental 
works, where the contribution of $\rho(1450)$ was taken into account~\cite{SNDetg2007,CMDetg2005}, 
the phase was fixed at $\varphi_{\rho^{\prime}} = 0^{\circ}$.

The results of the fit are used to calculate the radiative correction and the energy
spread correction. 
The difference between the corrections obtained with different fit solutions is considerable 
only for two energy points with $E = 1049.992$ MeV and $1060.206$ MeV. For them,
this difference is 1.5\% and 4\%, respectively, and is taken as systematic uncertainty.
The value of the Born cross 
section at a given energy point is determined as follows
\begin{equation}
\label{viscrs0}
\sigma = \frac{N_{sig}}{\varepsilon_{0} IL (1+\delta) (1+\delta_{E}) B(\eta\to 2\gamma)}.
\end{equation}
The values of $(1+\delta)$, $(1+\delta_{E})$ and the Born cross sections with statistical 
and systematic errors are listed in Table~\ref{tab2}. The main contributions to the systematic uncertainty
come from the uncertainties in the luminosity
measurement (0.6\%), the condition on the polar angle of photons (0.6\%), 
the condition on $\chi^2_{3\gamma}$ of the kinematic reconstruction (0.3-0.6\%),
the accuracy of $B(\eta\to 2\gamma)$ (0.5\%) and the energy spread correction.
The latter is due to the uncertainty in the measurement of $\sigma_E$ and reaches 0.7\%
for the energy point with $E = 1019.073$ MeV.
At the tails of the $\phi(1020)$ resonance the systematic uncertainty is dominated by 
the uncertainty of the number of signal events (up to 15\%).
The total systematic uncertainty is obtained as a sum in quadrature of all 
systematic uncertainties from different sources. 
It varies from 1\% to 15\% depending on the energy point
and is comparable to the statistical error at the vicinity of the $\phi(1020)$ resonance maximum.

The obtained energy scale shift for the 2024 scan $\Delta E_{24}$ (Table~\ref{FitTab})
significantly exceeds its systematic uncertainty (60 keV) estimated in Ref.~\cite{compton}. Although the origin of such a shift
is unclear, it does not affect the measured cross section.

The obtained values of the Born cross sections are actually numerical solutions
of the integral equation (\ref{viscrs}), in which the theoretical model is used
for regularization. Since the theoretical model is the same for all energy points, 
the Born cross sections obtained for different energy points can be correlated.
To study these correlations, a series of $10^4$ pseudoexperiments is used.
The number of signal events for each energy point is generated according to the theoretical model 
with a variance equal to its statistical error (Table~\ref{tab2}) with an additional term corresponding to~(\ref{desys})
added in quadrature.
The statistical uncertainties of
the Born cross section in Table~\ref{tab2} and in Fig.~\ref{crs_fit} correspond 
to the diagonal elements of the obtained covariance matrix. The full correlation matrix 
is given in the supplementary material.

\section{Discussion}

Figure~\ref{crs_diff} shows the relative differences between the measured 
$e^+e^- \to \eta\gamma$ Born cross section and the result of the fit described above for
this work and data from Ref.~\cite{SNDetg2000,CMDetg2001,SNDetg2007}.
The solid line on each plot shows the average difference, which is
($-4.5 \pm 1.2 \pm 2.7$)\% for the SND measurement using the decay channel $\eta\to3\pi^0$ in Ref.~\cite{SNDetg2007},
($-9.3\pm0.9\pm4.1$)\% for the CMD-2 measurement in Ref.~\cite{CMDetg2001} and ($-5.2 \pm 0.8 \pm 2.7$)\% for the SND measurement in Ref.~\cite{SNDetg2000}.
The first quoted error is statistical and the second is systematic. It is seen that 
the measurement made in this work has better accuracy and is higher compared to other measurements.

\begin{figure}[hb]
\center
\includegraphics[width=1.1\columnwidth]{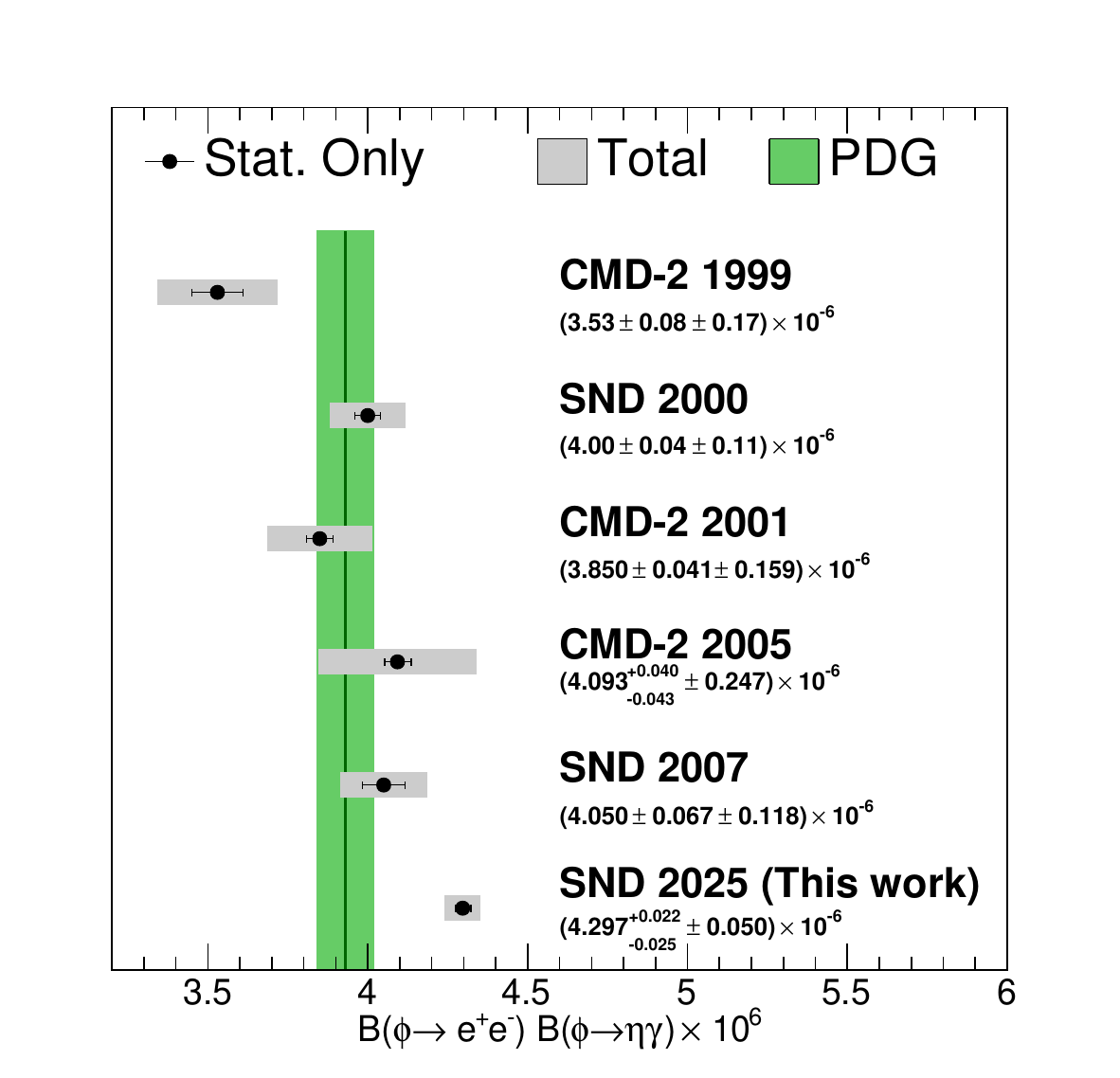}
\caption{\label{branching}
The value of the branching fractions product
$B(\phi\to e^+e^-)B(\phi\to \eta\gamma)$ obtained in this work for solution 1 compared to 
the world average value reported by PDG~\cite{pdg}
and to the measurements SND 2007~\cite{SNDetg2007}, CMD-2 2005~\cite{CMDetg2005},
CMD-2 2001~\cite{CMDetg2001}, SND 2000~\cite{SNDetg2000} and CMD-2 1999~\cite{CMDetg1999}.
}
\end{figure}

Using the relation $\sigma_{\phi} = 12\pi B_{\phi} / m_{\phi}^2$, where 
$B_{\phi} \equiv B(\phi\to e^+e^-)B(\phi\to \eta\gamma)$ is
the product of the branching fractions of the decays $\phi\to e^+e^-$ and $\phi\to\eta\gamma$
and $m_{\phi}$ is the $\phi$ meson mass~\cite{pdg}, we can obtain
\begin{eqnarray*}
B_{\phi} &=& (4.297^{+0.022}_{-0.025} \pm 0.050)\cdot 10^{-6}~~~\text{(solution 1)}, \nonumber \\
B_{\phi} &=& (3.993^{+0.024}_{-0.021} \pm 0.047)\cdot 10^{-6}~~~\text{(solution 2)}, \nonumber \\
B_{\phi} &=& (4.053^{+0.034}_{-0.028} \pm 0.052)\cdot 10^{-6}~~~\text{(solution 3)}, \nonumber \\
B_{\phi} &=& (4.332^{+0.024}_{-0.027} \pm 0.050)\cdot 10^{-6}~~~\text{(solution 4)}. \nonumber \\
\end{eqnarray*}

The first error is statistical and the second is systematic. 
The systematic uncertainty varies depending on the chosen solution.
The dominant contribution to it comes from the 
uncertainty of the measured cross section (1.1-1.2\%).
Variation of the parameters $\sigma_{\rho}$ and $\Gamma_{\phi}$ within their errors 
is used to estimate corresponding systematic uncertainty, which are 0.4-0.6\% and 0.2\%, respectively.
Since the obtained values for the parameters $\Delta E_{18}$,
$\sigma_{\rho^{\prime}}$, $\varphi_{\rho^{\prime}}$, $\sigma_{\phi^{\prime}}$
are largely determined by the data from Ref.\cite{SNDkskl,SNDetg23}, their contribution to the 
total error (0.1-0.3\%) is included into the systematic error. 
The total systematic uncertainty is obtained as a sum in quadrature of all 
systematic uncertainties from different sources.

Comparison of the obtained product $B(\phi\to e^+e^-)B(\phi\to \eta\gamma)$ with the previous measurements
makes sense only for the first solution,
since the interference pattern between the resonances for other solutions is 
significantly different from that assumed in previous experimental works.
Such a comparison is shown in Fig.~\ref{branching}.
Our measurement is higher than all previous ones in accordance with the differences of the measured
cross sections discussed above.

\section{Conclusion}

In the SND experiment at the VEPP-2000 collider, the measurement of the $e^+e^-\to\eta\gamma$ 
cross section has been performed in the center-of-mass energy range from 980 to 1060 MeV.
Our measurement is the most accurate to date and is higher compared to other measurements.
The total error of the measured cross section at the maximum of the $\phi(1020)$ 
resonance is $1.5$\%. From the fit to the cross section data with the vector meson dominance model, 
the value of the product 
$B(\phi\to e^+e^-)B(\phi\to \eta\gamma)$ has been obtained.

\end{document}